\documentstyle[epsf,12pt]{article}
\newcommand \beq{\begin{eqnarray}}
\newcommand \eeq{\end{eqnarray}}

\newcommand \la{\raisebox{-.5ex}{$\stackrel{<}{\sim}$}}
\newcommand \e{\varepsilon}

\newcommand \vp{{\bf v}_{\bf p}}

\newcommand{\rep}[1]{(\ref{#1})}

\newcommand{\yourtitle}[1]{
\mbox{}\\
\vskip 4\baselineskip
{\bf\noindent #1}\\ }

\newcommand{\youraddress}[1]{
\noindent\mbox{}\hspace{1in}\parbox[t]{4.0in}{#1}\\ }

\newcommand{\yournames}[1]{
\mbox{}\\
\mbox{}\\
\noindent\mbox{}\hspace{1in}{#1}\\ }

\newcommand{\yourabstract}[1]{
\mbox{}\\
\mbox{}\\
{\bf\noindent Abstract}\\
\begin{center}
\mbox{}\parbox[t]{5.in}{#1}
\end{center} }

\newcommand{\yoursection}[1]{
\vskip 2\baselineskip
{\bf\noindent #1}\\
\mbox{}\\
\vspace{-0.19in}}

\newcommand{\referencestyle}{
\small
\abovedisplayskip=6pt
\belowdisplayskip=6pt
\vspace{12pt}}

\begin{document}
\yourtitle{TRANSPORT PROPERTIES OF QUARK AND GLUON PLASMAS}
\yournames{H. Heiselberg}
\youraddress{Nuclear Science Division, MS 70A-3307\\
        Lawrence Berkeley Laboratory, \\
        University of California, Berkeley, California 94720}
\yourabstract{
The kinetic properties of relativistic quark-gluon and electron-photon
plasmas are described in the weak coupling limit.  The troublesome
Rutherford divergence at small scattering angles is screened by Debye
screening for the longitudinal or electric part of the interactions. The
transverse or magnetic part of the interactions is effectively screened
by Landau damping of the virtual photons and gluons transferred in the
QED and QCD interactions respectively. Including screening a number of
transport coefficients for QCD and QED plasmas can be calculated to
leading order in the interaction strength, including rates of momentum
and thermal relaxation, electrical conductivity, viscosities, flavor
and spin diffusion of both high temperature and degenerate plasmas.
Damping of quarks and gluons as well as color diffusion
in quark-gluon plasmas is, however, shown not to be sufficiently screened
and the rates depends on an infrared cut-off of order the ``magnetic
mass", $m_{\rm mag}\sim g^2 T$. }

\section{Introduction}

QCD plasmas consisting of quarks, antiquarks and gluons appear in a
number of situations. In the early universe the matter consisted mainly
of a quark-gluon plasma for the first microsecond before hadronization
set in. Present day
experiments at the CERN SPS, the Brookhaven AGS and future RHIC collider
are searching for the quark-gluon plasma in relativistic heavy ion
collisions.  Cold degenerate plasmas, $T\ll\mu$, of quarks may exist
in cores of neutron stars or in strangelets.

Relativistic QED and QCD plasmas have a lot of common features in the
weak coupling limit such as the Rutherford divergence in the elastic
differential cross section, the screening properties, and therefore also
transport properties.
In a plasma the typical momentum transfers
are of order $T$ or $\mu$, whichever is the larger.
For sufficiently high temperature or density, the running coupling
constant $\alpha_s(Q)$ is small and
we can treat the quark-gluon plasma as weakly interacting. In fact lattice
gauge calculations indicate that the quark-gluon plasma behaves much like
a free
gas already at temperatures not far above the critical temperature,
$T_c\simeq 160$ MeV, at which the phase transition between hadronic matter and
quark-gluon plasma takes place. Anyway, since we are better at interpolating
than extrapolating, knowledge of the behavior of weakly interacting
quark-gluon plasmas at high temperatures or densities should be very useful.

We emphasize that it is the effect of Landau damping which
effectively leads to screening of transverse interactions
and give the characteristic
relaxation rates in transport processes and some transport coefficients
for weakly interacting electron-photon and quark-gluon plasmas for both
thermal plasmas \cite{BP1,BP2} as well as degenerate ones
\cite{deg}. In addition, we discuss
the quark and gluon quasiparticle damping rates and
the rates of color diffusion in which
the transverse interactions are not sufficiently
screened so that an infrared cut-off on the order
of the ``magnetic mass", $m_{\rm mag}\simeq g^2T$, is needed.

\section{Screening in QCD and QED}
The static chromodynamic interaction between two quarks
in vacuum gives the matrix element for near-forward elastic scattering
\beq
    M(q) = \frac{\alpha_s}{q^2}\left( 1 -
   \frac{{\bf v}_1\times{\bf\hat{q}}}{c}
   \frac{{\bf v}_2\times{\bf\hat{q}}}{c} \right)  \, , \label{M}
\eeq
where $\alpha_s=g^2/4\pi$ is the QCD fine structure constant,
${\bf v}_1$ and ${\bf v}_2$ are the velocities of the two
interacting particles
and ${\bf q}$ is the momentum transfer in the collision.
The first part is the electric or
longitudinal (+timelike) part of the interactions. The second part is the
magnetic or transverse part of the interactions.
In the Born approximation the corresponding potential is obtained by
Fourier transform of \rep{M} by which one obtains the standard
Coulomb and Lorenzt interactions respectively.
A weakly interacting QCD plasma is very similar to a QED plasma  if one
substitutes the fine structure constant $\alpha_s$
by  $\alpha=e^2\simeq 1/137$,
the gluons by photons and the quarks by leptons with
the associated statistical factors. The interaction via the gluon
field is determined by gauge symmetry in much the same way as in QED and
therefore scattering by a gluon exchange is very similar to
that by photon exchange and the Feynman diagrams carry over. There is
one crucial difference, namely that the gluon can couple directly to
itself. This leads to confinement and the running coupling constant,
$\alpha_s(Q)$. For a given coupling constant, however, the kinetic
properties are very similar as we shall see.

In non-relativistic plasmas such as the electron plasma in terrestrial
metals, the electron  velocities are of order the Fermi velocity, which
is much smaller than the speed of light, and therefore the transverse
interactions are usually ignored. In relativistic plasmas they are,
however, of similar magnitude and for degenerate plasmas the transverse
interactions may in fact dominate the transport properties, as will be
described below, because they are effectively less screened than the
longitudinal interactions.

By squaring the matrix element we obtain the Rutherford differential
cross section at small momentum transfer ${\bf q}$
in the center-of-mass system
\beq
    \frac{d\sigma}{d\Omega} = 4E^2\frac{\alpha_s^2}{q^4}(1+v^2)
            \, . \label{Ru}
\eeq
The momentum transfer is related to the scattering angle
$\theta$ by $q=2E\sin(\theta/2)$ where $E$ is the particle energy in
the center-of-mass system. We observe that \rep{Ru}
diverges as $\theta^{-4}$ at small angles and the total cross
section is infinite signifying a long-range interaction.
In calculating transport properties one typically weights the differential
cross section by a factor $q^2=2E^2(1-\cos\theta)$ but that still leads
to a logarithmically diverging integral and therefore to vanishing transport
coefficients.

In a medium this singularity is screened as given by the Dyson equation
in which a gluon self energy $\Pi_{L,T}$ is added to the propagator
$t^{-1}=\omega^2-q^2$. For the matrix element this gives
\beq
    t^{-1} \to \omega^2-q^2-\Pi_{L,T} \, , \label{Dyson}
\eeq
(we refer to Weldon \cite{We} for details on separating longitudinal
and transverse parts of the interaction)
where the longitudinal and transverse parts of the
selfenergy in QED and QCD are for $\omega, q\ll T$ given by
\beq
  \Pi_L(\omega,q) &=& q_D^2 \left(1-\frac{x}{2}\ln\frac{x+1}{x-1}\right)
  \,  ,\label{PiL} \\
  \Pi_T(\omega,q)  &=& q_D^2 \left[\frac{1}{2}x^2+\frac{1}{4}x(1-x^2)
      \ln\frac{x+1}{x-1}\right] \, , \label{PiT}
\eeq
where $x=\omega/qv_p$ and $v_p=c$ for the relativistic plasmas considered
here.  The Debye screening wavenumbers in QCD is
\beq
   q_D^2 &=& g^2 \displaystyle{
         \left((N_q+2N)\frac{T^2}{6} + N_q\frac{\mu_q^2}{2\pi^2} \right) }
          \label{qD} \, ,
\eeq
where $N=3$ is the number of colors, $N_q$ is the number of quark flavors,
$T$ the plasma temperature and $\mu_q$ the quark chemical potential.
We refer to \cite{LH} for a detailed comparison to QED plasmas.

In the static limit, $\Pi_L(\omega=0,q)=q_D^2$, and the longitudinal
interactions are Debye screened. Consequently, the typical elastic scattering
and transport cross sections due to longitudinal interactions alone become
\beq
   \sigma^L_{el} &=& \int \frac{d\sigma}{d\Omega} d\Omega
            \simeq \alpha_s^2 \int \frac{d^2q}{(q^2+q_D^2)^2}
         \simeq \frac{\alpha_s}{T^2}
       \, . \label{sL} \\
   \sigma^L_{tr} &\simeq& \int \frac{d\sigma}{d\Omega}(1-\cos\theta) d\Omega
          \simeq  \frac{\alpha_s^2}{E^2}\int  \frac{q^2d^2q}{(q^2+q_D^2)^2}
          \simeq \frac{\alpha_s^2}{E^2}\ln(T/q_D)  \, . \label{strL}
\eeq
in a high temperature quark-gluon plasma, where particles energies
are $E\sim T$.

For the transverse interactions the selfenergy
obeys the transversality condition
$q^\mu \Pi_{\mu\nu} = 0$,
which insures that the magnetic interactions are
unscreened in the static limit, $\Pi_T(\omega=0,q)=0$.
Consequently, the cross sections corresponding to \rep{sL} and \rep{strL}
diverge leading to zero transport coefficients.
It has therefore been suggested that the transverse interactions are cut
off below the ``magnetic mass", $m_{mag}\sim g^2T$, where infrared
divergences appear in the plasma \cite{Linde}.
However, as was shown in \cite{BP1} and as will be shown below,
dynamical screening due to Landau damping effectively screen the
transverse interactions off in a number of situations
at a length scale of order the Debye screening
length $\sim 1/gT$ as in Debye screening.
Nevertheless, there are three important length scales in the
quark-gluon plasma. For a hot plasma they are, in increasing size,
the interparticle spacing
$\sim 1/T$, the Debye screening length $\sim 1/gT$, and the
scale $1/m_{mag}\sim 1/g^2T$ where QCD effects come into play.

\section{Transport Coefficients for Hot Plasmas}

In this section we calculate a number of transport coefficients for
weakly-interacting, high temperature plasmas.  We work in the kinematic
limit  of massless particles, $m\ll T$, and zero chemical potential,
$|\mu\ll T|$.

     The characteristic timescales, $\tau$, describing the rate at which a
plasma tends towards equilibrium if it is initially produced out of
equilibrium, as in a scattering process, or if driven by an external field,
are determined by solving the kinetic equation.  For a system with well
defined quasiparticles, the Boltzmann transport equation is
\beq
   ( \frac{\partial}{\partial t} + {\bf \vp}\cdot\nabla_{\bf r} &+&
   {\bf F}\cdot\nabla_{\bf p} ) n_{\bf p}
    =\,  - 2\pi\nu_2\sum_{234} |M_{12\to 34}|^2
    \delta_{{\bf p}_1+{\bf p}_2 ,{\bf p}_3+{\bf p}_4}
    \delta (\varepsilon_1 +\varepsilon_2 -\varepsilon_3 -\varepsilon_4 )
    \nonumber \\
    &\times&[ n_{{\bf p}_1}n_{{\bf p}_2}(1\pm n_{{\bf p}_3})
    (1\pm n_{{\bf p}_4})  - n_{{\bf p}_3}
    n_{{\bf p}_4}(1\pm n_{{\bf p}_1})(1\pm n_{{\bf p}_2})] \, ,\label{BE}
\eeq
where ${\bf p}$ is the quasiparticle momentum, $n_{\bf p}$ the
quasiparticle distribution function and ${\bf F}$ the force on a
quasiparticle. The r.h.s. is the collision integral
for scattering particles from initial states 1 and 2 to
final states 3 and 4, respectively.
The $(1\pm n_p)$ factors correspond physically to the Pauli blocking of
final states, in the case of fermions, and to (induced or) stimulated
emission, in the case of bosons.

The matrix element for the scattering process $1+2\to 3+4$ is
 $|M_{12\to34}|^2 = |{\cal M}_{12\to34}|^2/16\e_1\e_2\e_3\e_4$
where ${\cal M}$ is
the Lorentz invariant matrix element normalized in the usual manner in
relativistic theories. When electrons, quarks and
gluons scatter elastically through photon or gluon exchange in a
vacuum, the matrix element squared averaged over initial
and summed over final states is
dominated by a $t^2=(\omega^2-q^2)^2$ singularity, for example for
quark-gluon scattering
\beq
    |{\cal M}_{qg\to qg}|^2 =g^4(u^2+s^2)/t^2 .  \label{Mqg}
\eeq
The gluon-gluon and quark-quark matrix elements only differ by a factor
9/4 and 4/9 respectively near small momentum scattering.

\subsection{Viscosities}

In Ref. \cite{BP1,LH}  the first viscosity of a quark-gluon plasma was
derived to leading logarithmic order in the QCD coupling
strength by solving the Boltzmann kinetic equation. For gluons
\beq
    \eta_g = \frac{2^9 15\xi(5)^2}{\pi^5(1+11N_q/48)}
   \frac{T^3}{g^4\ln(T/q_D)}
     = 0.34 \frac{T^3}{\alpha_s^2\ln(1/\alpha_s)} \, . \label{etag}
\eeq
Second, the quarks carry momentum, and therefore produce an increase in
the total viscosity, $\eta=\eta_g+\eta_q$.
The quark viscosity, $\eta_q$, is obtained analogously to the gluon one
\beq
   \eta_q =
   2.2 \frac{1+11N_q/48}{1+7N_q/33} N_q \eta_g \, , \label{etaq}
\eeq
which for $N_q=2$ results in $\eta_q=4.4\eta_g$, a quark viscosity that
is larger than the gluon one because the gluons generally
interact stronger than the quarks.

Writing each $\eta_i$ $(i=q,g)$ in terms of the viscous
relaxation time, $\tau_{\eta, i}$, as
\beq
    \eta_i = w_i\tau_{\eta, i}/5,   \label{etai}
\eeq
where $w$ is the enthalpy,
we obtain the viscous relaxation rate for gluons
\beq
    \frac{1}{\tau_{\eta,g}} = 4.11
    (1+{11{N_q}\over 48})T\alpha_s^2\ln(1/\alpha_s),  \label{tg}
\eeq
and for quarks and antiquarks
\beq
     \frac{1}{\tau_{\eta,q}} = 1.27 (1+\frac{7N_q}{33})
     T \alpha_s^2\ln(1/\alpha_s) \, .  \label{tq}
\eeq

The second viscosity $\zeta$ is zero for a gas of massless relativistic
particles.
 Thermal conduction is not a hydrodynamic mode in relativistic plasmas
with zero chemical potential.

\subsection{Momentum Relaxation Times}

In \cite{BP2} the time for
transfer of momentum between two interpenetrating, spatially uniform
plasmas in relative motion  has been calculated.
Elastic collisions between the two
interpenetrating plasmas lead to a relative flow velocity decreasing as
function of time and the characteristic stopping time is the ``momentum
relaxation time".
  We emphasize that we are only considering collisional phenomena
in this  calculation. It might be the case that collective phenomena,
such as the two-stream instability could lead to relaxation
faster than that due to collisions.

 The resulting momentum relaxation rate
$\tau^{-1}_{{\rm mom},qg}$ for gluons colliding with quarks and antiquarks
($\nu_2=12N_q$), exact to leading logarithmic order in
$\alpha_s=g^2/4\pi$, is
\beq
    1/\tau_{\rm mom,qg} = \frac{20\pi}{7}(1+\frac{21}{32}N_q)
    \alpha_s^2 \ln(1/\alpha_s) T  \, . \label{tmom}
\eeq
Momentum relaxation rates for two plasmas with different quark flavors,
spins, or colors, or for different gluon colors or spins have the same form,
$\sim \alpha_s^2\ln(1/\alpha_s) T$.

 In QED plasma similar stopping times are obtained
for lepton stopping on antileptons \cite{BP2,LH} as in \rep{tmom} when
$\alpha_s$ is replaced by $\alpha$. Photon stopping rates are, however,
different, $\tau_{\rm mom,\gamma\ell}\simeq \alpha_s^2 T$, because the
photon does not couple to itself as does the gluon and so it does not
interact with a lepton by exchanging a photon but only through
Compton scattering.

\subsection{Electrical Conductivity}

    Another transport coefficient of interest is the electrical conductivity,
$\sigma_{el}$, of the early universe.  The principal conduction process is
flow of charged leptons, and the dominant scattering process is
electromagnetic interaction with other charged
particles; strongly interacting particles have much shorter mean-free paths.
The infrared singularity of the transverse interaction in QED is treated as in
QCD, only now $q_D^2=N_l e^2 T^2 /3$, where $N_l$ is the number of charged
lepton species present at temperature $T$.  To calculate $\sigma_{el}$ we
consider the current of charged leptons (1) and antileptons (2) in a static
electric field, ${\bf E}$.  Taking the components to be thermally distributed
with opposite fluid velocities, ${\bf u{}_1=-{\bf u}_2}$,
the total electrical current
is ${\bf j}_{\ell\bar{\ell}} = -e n_{\ell\bar{\ell}}{\bf u}_1$, where the
density of electric charge carriers is $n_{\ell\bar{\ell}}=3\zeta(3)N_\ell
T^3/ \pi^2$.  Solving the kinetic equation \rep{BE} we find the electrical
conductivity
\beq
    \sigma_{\ell\bar{\ell}} = j_{\ell\bar{\ell}}/E =
   \frac{3\zeta(3)}{\ln2} \frac{1}{\alpha\ln(1/\alpha)} T\, .\label{sll}
\eeq
The related electric
current relaxation time is
\beq
     \frac{1}{\tau_{\ell\bar{\ell}}} = \frac{4\pi N_{\ell
      }\ln2}{27\zeta(3)}\alpha^2\ln{(1/\alpha)} T , \label{tll}
\eeq
which is very similar to the corresponding momentum relaxation time
\rep{tmom} in QED.

Although quarks (of charge $Q_qe$) contribute negligibly to the electrical
current, their presence leads to additional stopping of the leptons and thus
smaller conductivity.  Adding contributions from $\ell q$ and $\ell\bar q$
collisions we obtain the total electrical conductivity $ \sigma_{el}=
\sigma_{\ell\bar{\ell}} /(1+3\sum_{q=1}^{N_q} Q^2_q) $.

\section{Transport Coefficients in Degenerate Matter}

In degenerate QED and QCD plasmas there are practically no photons
or gluons present respectively  since $T\ll\mu$. The Debye screening
length is according to Eq. \rep{qD} proportional to $\lambda_D\simeq
1/e\mu$ for an electron plasma and $\lambda_D\simeq 1/g\mu$ for a quark
plasma.

In a degenerate plasma there are three momentum scales, namely
$\mu$, $T$, and $q_D$, whereas in the hot plasmas we could neglect the
chemical potential. In the cold plasma $T\ll\mu$ and likewise for the
weakly interacting plasma $q_D\ll\mu$, but it is important to consider
the limits of $q_D\ll T$ and $T\ll q_D$ separately.

Momentum transfer processes in degenerate quark matter, $T\ll\mu$
(chemical potential), as for example in neutron stars, are also
characterized by the rate of momentum relaxation for strong
interactions, $\tau_{{\rm mom}}^{-1}$. For $N_q$ quark flavors with the
same chemical potentials we find, neglecting the strange quark mass,
that \cite{deg}
\beq
   \frac{1}{\tau_{{\rm mom}}}&=& \frac{8N_q}{3\pi}T\alpha_s^2
   \left\{ \begin{array}{ll}
   (3/2)\ln(T/q_D) , & \, T\gg q_D  \\
   a(T/q_D)^{2/3} + (\pi^3/12)(T/q_D), & \, T\ll q_D
   \end{array} \right\},     \label{tdeg}
\eeq
where $a=(2\pi)^{2/3}\Gamma(8/3)\zeta(5/3)/6\simeq 1.81$.

The two limiting cases can be qualitatively understood by noticing that,
due to Pauli blocking and energy conservation, the energy of the
particles before and after collisions must be near the Fermi
surface, and therefore $\omega\la T$. For $T\gg q_D$
the limitation $\omega\la T$ does not affect the
screening because $|\omega|\le q\sim q_D\ll T$ and   consequently the
result \rep{tdeg} is analogous to that for  high temperatures, Eq.
\rep{tg}, but with $q_D^2=2N_q\alpha_s\mu^2/\pi$. The result for
$T\ll q_D$ is qualitatively different due to Landau damping of modes
with $q\la (\omega q_D^2)^{1/3}$, where now $\omega\sim T$.
The two terms in \rep{tdeg} correspond to the
contributions from transverse and longitudinal interactions respectively
and we note that transverse interactions dominate for $T\ll q_D$ because
Landau damping is less effective than Debye screening in screening
interactions at small $q$ and $\omega$.

The viscous relaxation time for quarks is defined analogously to
\rep{etai} by
\begin{eqnarray}
  \eta= \frac{1}{5} np_Fv_F \tau_\eta   \, ,
\end{eqnarray}
where $p_F=\mu_q$ is the Fermi momentum, $v_F=c=1$ is the Fermi velocity
and $n=N_qn_q$ is the density of quarks. By solving the kinetic
equation we find \cite{deg}
\begin{eqnarray}
   \frac{1}{\tau_\eta} &=& \frac{8}{\pi}N_q\alpha_s^2  \times
    \left\{ \begin{array}{ll}
   \displaystyle{\frac{5}{3}}T\ln(T/q_D) \, , & \, T\gg q_D \\
  \displaystyle{a\frac{T^{5/3}}{q_D^{2/3}}
   +\frac{\pi^3}{4} \frac{T^2}{q_D} } \, ,  &  T\ll q_D
  \end{array} \right\}  \, .     \label{teta}
\end{eqnarray}

We define a characteristic  relaxation time for thermal conduction,
$\tau_\kappa$, by
\begin{eqnarray}
  \kappa= \frac{1}{3}v_F^2 c_v \tau_\kappa \, , \label{ka}
\end{eqnarray}
where $c_v=(N_q/2)T\mu_q^2$ is the specific heat per volume.
Thus we find \cite{deg}
\begin{eqnarray}
   \frac{1}{\tau_\kappa} &=&
   \frac{4}{\pi^3}N_q\alpha_s^2 \mu_q^2
   \times \left\{ \begin{array}{ll}
   \displaystyle{\frac{\ln(T/q_D)}{3T} } \, ,
              &  T\gg q_D \\
              &\\
   \displaystyle{2\zeta(3) T/q_D^2
   }    \, ,  &  T\ll q_D
   \end{array} \right\}   \,   .   \label{tkappa}
\end{eqnarray}
The relaxation time for thermal conduction  has a different temperature
dependence than both $\tau_s$ and $\tau_\eta$  because the thermal
conduction was weighted by energy transfers ($\omega^2$) instead of
momentum transfers ($q^2$) as for momentum stopping, electrical
conduction and viscous processes.

Applications to burning of nuclear
matter into strange quark matter in the interior of a neutron star
are described in Ref. \cite{deg}.
In the above calculation the quark matter was assumed to be present in bulk.
The transport properties may, however, be significantly different
in a complex mixed phase of quark and nuclear matter in cores of neutron
stars \cite{Droplet}.

\section{Damping of Quarks and Gluons\hfil}

Historically, calculating damping rates of  quasiparticles in a thermal
quark-gluon plasma is a controversial subject, since early results
indicated that the damping rate was negative, and gauge dependent
(for a review, see Ref. \cite{Pisarski}).  Here
we show how the lifetime may be calculated within the framework of the
kinetic equation \cite{HP}.
The quasiparticle decay rate for a gluon of momentum
${\bf p}_1$ scattering on other gluons is \cite{BPbook}
\beq
    \frac{1}{\tau_{p_1}^{gg}} &=& 2\pi\nu_2\sum_{{\bf q,p}_2}
    \frac{n_2(1+n_3)(1+n_4)}{1+n_1} |M_{gg\to gg}|^2
    \delta (\varepsilon_1 +\varepsilon_2 -\varepsilon_3 -\varepsilon_4 )
     \, , \label{qlife}
\eeq
where ${\bf p}_2$ is the momentum of the other gluon.
When $p_1\gg q_D\sim gT$ the integrals in \rep{qlife} over the transverse
part of the interaction diverge as
shown in \cite{HP} because the factor $(1\cos\theta)$ or $q^2$ appearing in
the earlier transport calculations is missing;
i.e., Landau damping alone is
insufficient  to obtain a non-zero quasiparticle lifetime.
Including an infrared cut-off, $\lambda\simeq m_{mag}\simeq g^2T$, which takes
into account the failure of perturbative ideas at momentum scales of order
the magnetic mass,
the leading contribution comes from small momentum transfers $q\sim q_D\sim gT$
and we obtain
\beq
    1/\tau_{p_1}^{(g)} &=&   3\alpha_s \ln(1/\alpha_s) \, T  \, . \label{q4}
\eeq
The quasiparticle decay rate for quarks can be calculated analogously and
is just 4/9 of that for the gluon, the factor coming from the matrix elements
at small momentum transfer
for quark scattering on quarks and gluons compared with those for gluon
scattering.

Let us now compare this result with that obtained using field-theoretic
techniques.
Braaten \& Pisarski \cite{Pisarski} have developed a technique for resumming
soft thermal loops which provides screening so that the damping,
$\gamma$, at $p_1=0$ is positive and gauge independent. Recently, Burgess,
and Marini \& Rebhan \cite{Rebhan} have obtained the leading logarithmic
order for quarks and gluons with momenta $p_1\gg gT$. They evaluate the
gluon self energy, given by the gluon bubble, to leading order by
including screening in the propagator of the soft gluon in the bubble
and introducing the same cutoff. Their result for the imaginary part of
the self energy, which is one half the quasiparticle decay rate, agrees
with ours, Eq. \rep{q4}, because exchange contributions (vertex
corrections),  which are automatically included in the kinetic equation,
do not contribute to leading order.

     The relaxation rates in transport processes \cite{BP1} are of order
$\sim \alpha_s^2 T$, i.e., suppressed by a factor $\alpha_s$ with respect to
the damping rates. This is because in transport processes one has
an extra factor $q^2$ in integrals like \rep{BE}
which suppresses the rate by a factor $q_D^2/T^2\simeq \alpha_s$.

\section{Flavor, Spin and Color Diffusion}

 Let us first consider the case where the particle flavors have
been separated spatially, i.e.,
the flavor chemical potential depends on position, $\mu_i({\bf r})$.
In a steady state scenario the flavor will then be flowing with
flow velocity, $u_i$. If we make the standard ansatz for the
distribution function (see, e.g., \cite{BPbook,HP})
\beq
   n_{{\bf p},i}=\left(\exp(\frac{\epsilon_{\bf p}-\mu_i({\bf r})-
                  {\bf u}_i\cdot{\bf p}}{T})+1\right)^{-1}
     \simeq \left(\exp(\frac{\epsilon_{\bf p}-
            \mu_i({\bf r})}{T})+1\right)^{-1} -
   \frac{\partial n_p}{\partial\epsilon_p} {\bf u}_i\cdot{\bf p} \, ,
   \label{ni}
\eeq
the expansion is valid near global equilibrium where $\mu_i$ and
therefore also ${\bf u}_i$ is
small. The two terms are those of local isotropic equilibrium and the
deviation from that. In general the deviation from local equilibrium has
to be found selfconsistently by solving the Boltzmann equation. However,
in the case of the viscosity the analogous ansatz was found to be very
good \cite{BP1}. The flavor current is then simply
  $ {\bf j}_i = \sum_{\bf p} n_{{\bf p},i} = {\bf u}_i n_i $
where $n_i=\sum_p n_{p,i}$ is the density of a particular flavor {\it i}.
Linearizing the Boltzmann equation we now find
\beq
    -\frac{\partial n_1}{\partial\epsilon_1} {\bf v}_1\cdot\nabla\mu_i
    &=& 2\pi\sum_{234} n_1n_2(1-n_3)(1-n_4)
    \delta_{{\bf p}_1+{\bf p}_2 ,{\bf p}_3+{\bf p}_4}
    \delta (\varepsilon_1 +\varepsilon_2 -\varepsilon_3 -\varepsilon_4 )
    \nonumber \\
    &\times&  |M_{12\to 34}|^2 ({\bf u}_1-{\bf u}_2)\cdot {\bf q }
\eeq
The resulting flavor diffusion coefficient defined by:
  $  {\bf j}_i = -D_{flavor} \nabla\mu_i  $,
is now straightforward to evaluate when the screening is properly
included as described above. The calculation is analogous to
that of the momentum stopping or viscous times. We find
\beq
    D_{flavor}^{-1} \simeq  2.0(1+\frac{N_f}{6})
        \alpha_s^2\ln(1/\alpha_s) T \, .\label{Di}
\eeq

  Subsequently, let us consider the case where the particle spins have
been polarized spatially, i.e.,
the spin chemical potential depends on position, $\mu_\sigma({\bf r})$.
With the analogous ansatz to \rep{ni} for the distribution functions with
$\mu_\sigma$ instead of $\mu_i$
we find the spin current ${\bf j}_\sigma={\bf u}_\sigma n_\sigma$.
Linearizing the Boltzmann equation we find
\beq
    -\frac{\partial n_1}{\partial\epsilon_1} {\bf v}_1\cdot\nabla\mu_\sigma
    &=& 2\pi\sum_{234} n_1n_2(1-n_3)(1\pm n_4)
     \delta_{{\bf p}_1+{\bf p}_2 ,{\bf p}_3+{\bf p}_4}
    \delta (\varepsilon_1 +\varepsilon_2 -\varepsilon_3 -\varepsilon_4 )
    \nonumber \\
    &\times& \left[ |M_{12\to 34}^{\uparrow\uparrow}|^2
            ({\bf u}_1-{\bf u}_2)\cdot {\bf q}
           +|M_{12\to 34}^{\uparrow\downarrow}|^2
            ({\bf u}_1-{\bf u}_2)\cdot({\bf p}_1+{\bf p}_2) \right] \, ,
\eeq
where $M^{\uparrow\downarrow}$ and $M^{\uparrow\uparrow}$ refer to the
spin flip and the non-spin flip parts of the amplitude.

The transition current can be decomposed
into the non-spin flip and the spin flip parts
by the Gordon decomposition rule
\beq
    J_\mu &\simeq& g \bar{u}_f\gamma_\mu u_i
       \simeq   \frac{g}{2m} \bar{u}_f\left[
          (p_f+p_i)_\mu +i\sigma_{\mu\nu}(p_f-p_i)^\nu
          \right] u_i \, .
\eeq
We notice that
the latter is suppressed by a factor $q=p_f-p_i$ which leads to a
spin flip amplitude suppressed by a factor $q^2$. We then find that
the spin flip interactions do not contribute to collisions to leading
logarithmic order and the collision integral is similar to those evaluated
above. Consequently, the corresponding quark spin
diffuseness parameter is
\beq
    D_{\sigma}^{(q)} = D_{flavor}  \, .\label{Ds}
\eeq
Gluon diffusion is slower,
$D_{\sigma}^{(g)}\simeq (4/9)^2 D_{\sigma}^{(q)}$,
because they interact stronger.

 Finally, let us, like for the spin diffusion, assume that color has
been polarized
spatially given by a color chemical potential, $\mu_c({\bf r})$.
The basic difference to spin diffusion is that {\it quarks and gluons
can easily flip color directions} in forward scattering
by color exchanges, i.e., one does not pay the
extra $q^2$ penalty factor as in the case of
spin flip. Consequently, the color flip
interactions will dominate the collisions since they effectively reverse
the color currents.  As a consequence the collision term becomes similar
to that for the quasiparticle scatterings \rep{tg} and we find for
the color diffuseness parameter
\beq
    D_{color}^{-1} \simeq  0.3\, \alpha_s\ln(1/\alpha_s) T
             \sim \alpha_s^{-1} D_\sigma^{-1}  \, .\label{Dc}
\eeq
The color flip mechanism amplifies the collisions so the color cannot
diffuse as easily as spin or flavor.

The color conductivity is found analogous to the electrical
conductivity, where $\sigma_{el}\simeq q_D^2\tau_{el}$ (see Eqs. \rep{sll}
and \rep{tll}),
\beq
    \sigma_c \sim q_D^2 D_{color} \sim T/\ln(1/\alpha_s) \, .\label{sc}
\eeq
The characteristic relaxation times for conduction are very
different in QCD, where
$\tau_{color}\sim D_{color}\sim (\alpha_s\ln(1/\alpha_s)T)^{-1}$,
as compared to QED, where
$\tau_{el}=\tau_{mom}\simeq (\alpha^2\ln(1/\alpha)T)^{-1}$.
Consequently, QGP are much poorer color conductors than QED plasmas for
the same coupling constant.

These surprising results for QCD are qualitatively in agreement with those
found by Selikhov \& Gyulassy \cite{GS} who have considered the diffusion
of color in color space. They use the fluctuation-dissipation
theorem to estimate the deviations from equilibrium and find the same
two terms as in \rep{Dc}, which they denote the momentum and color diffusion
terms, and they also find that the latter dominates.
Inserting the same infrared cut-off they find a color diffusion
coefficient in color space equal to \rep{q4}
\beq
    d_c = \frac{1}{\tau^{(g)}_1}
        =  3\alpha_s\ln(1/\alpha_s) T \, . \label{dc}
\eeq
Note that this quantity is proportional to the inverse of $D_{color}$
of Eq. \rep{Dc}.

The color flip mechanism is not restricted to QCD but has analogues in
other non-abelian gauge
theories. In the very early universe when $T > T_{SSB}\simeq 250$ GeV, the
$W^\pm$ bosons are massless and faces the same
electroweak screening problems as QCD and QED. Since
now the exchanged $W^\pm$ bosons carry charge (unlike the photon, but
like the colored gluon), they can easily change the charge of, for example,
an electron to a neutrino in forward scatterings. Thus
the collision term will lack the usual factor $q^2$  as for the
quasiparticle damping rate and the color diffusion.
Since $SU(2)\times U(1)$ gauge fields should have the same
infrared problems as SU(3) at the scale of the magnetic mass, $\sim e^2 T$,
we insert this infrared cutoff. Thus we find a diffusion time
for charged electroweak particles in the very early universe of order
\beq
   \tau_{diff} \sim (\alpha \ln(1/\alpha) T)^{-1}  ,
\eeq
which  is a factor $\alpha$ smaller than the momentum stopping time.
The electrical conductivity will be smaller
by the same factor as well.

\section{Summary}

Whereas the troublesome
Rutherford divergence at small scattering angles is screened by Debye
screening for the longitudinal or electric part of the interactions, the
transverse or magnetic part of the interactions is effectively screened
by Landau damping of the virtual photons and gluons transferred in the
interactions. Including the screening, we calculated
a number of transport coefficients for QED and QCD plasmas to
leading order in the interaction strength. These included rates of momentum
and thermal relaxation, electrical conductivity, and viscosities of
quark-gluon plasmas for thermal as well as degenerate plasmas.

Our calculations above show that the transport properties of high
temperature and degenerate QCD and QED plasmas of relativistic  particles
have several interesting features. In a degenerate plasmas there are three
scales: the chemical potentials, the temperature, and the Debye
screening wavenumber, whereas in thermal plasmas, where $\mu\ll T$,
the chemical potentials can be neglected. In the degenerate case, where
$\mu$ is much larger than both $T$ and $q_D$, the physics changes
between the two limiting cases of $T\gg q_D$ and $T\ll q_D$. When $T\gg
q_D$ the characteristic  relaxation rates are $1/\tau\simeq
\alpha_s^2\ln(T/q_D)T$ as in the high temperature plasma, and the
contributions from longitudinal and transverse interactions are of
comparable magnitudes. However, the Debye wavenumber is different,
$q_D\sim gT$ at high temperatures and $q_D\sim g\mu$ in degenerate
plasmas. When $T\ll q_D$ the transverse interactions dominate the
scattering processes because they are screened by Landau damping only
for energy transfers less than $\sim(q_D^2T)^{1/3}$. The resulting
relaxation rates for momentum transfer, electrical conduction, and
viscous processes  scale as $1/\tau\sim (\alpha_sT)^{5/3}/\mu_q^{2/3}$,
while the  relaxation rate for thermal conduction is  $1/\tau_\kappa\sim
\alpha_sT$. The qualitative reason for $\tau_\kappa$ behaving in a
different way from the  other relaxation times is related to the
singular character of the interaction for small energy transfer and
small momentum transfer. An important general conclusion of these
studies of QCD and QED plasmas is that the the transport coefficients
deviate from the standard results of Fermi liquid theory and relaxations
times are different for different transport processes.

Color diffusion and the quark and gluon quasiparticle
      decay rates are not sufficiently screened and do depend
      on an infrared cut-off of order the magnetic mass, $m_{mag}\sim g^2T$;
typically $\tau^{-1}\sim \alpha_s\ln(q_D/m_{mag})T
                    \sim \alpha_s\ln(1/\alpha_s)T$.
As a consequence, quasiparticle decay is fast,
color diffusion is slow and the QGP is a poor color conductor.

\yoursection{Acknowledgments}

This work was supported in part by U. S. National Science Foundation
grants No. PHY89-21025 and No. PHY91-00283, NASA grant No. NAGW-1583,
DOE grant No. DE-AC03-76SF00098 and the Danish
Natural Science Research Council. Discussions with Gordon Baym and
Chris Pethick are gratefully acknowledged.

\end{document}